%% file: root.tex
\documentclass[preprint,12pt]{elsarticle}

\usepackage{acro}
\usepackage{hyperref}
\usepackage{amssymb}
\usepackage{amsmath}
\usepackage{amsthm}
\usepackage{graphicx}
\usepackage{multirow}

\theoremstyle{remark}

\usepackage{svg}
\usepackage{cleveref}
\usepackage{diagbox}
\input{acros}
\usepackage{todonotes}
\usepackage{booktabs}

\usepackage{enumitem}
\usepackage{comment}
\usepackage{xcolor}

\usepackage{microtype}
\newif\ifworkinprogress
\workinprogresstrue 
\ifworkinprogress

\else

\fi

\journal{Information Fusion}

\begin{document}

\begin{frontmatter}



    \title{Cross-Modal Binary Attention: An Energy-Efficient Fusion Framework for Audio-Visual Learning}


    \author[a1]{Mohamed Saleh} 
    \author[a1,a2]{Zahra Ahmadi\corref{cor1}} \ead{Ahmadi.Zahra@mh-hannover.de} 

    \affiliation[a1]{organization={Peter L. Reichertz Institute for Medical Informatics of TU Braunschweig and Hannover Medical School},
            city={Hannover},
            country={Germany}}
\cortext[cor1]{Corresponding author}
\affiliation[a2]{%
  organization={Lower Saxony Center for AI and Causal Methods in Medicine (CAIMed)},
  city={Hannover},
  country={Germany},
  }

    \begin{abstract}
        Effective multimodal fusion requires mechanisms that can capture complex cross-modal dependencies while remaining computationally scalable for real-world deployment. Existing audio-visual fusion approaches face a fundamental trade-off: attention-based methods effectively model cross-modal relationships but incur quadratic computational complexity that prevents hierarchical, multi-scale architectures, while efficient fusion strategies rely on simplistic concatenation that fails to extract complementary cross-modal information. We introduce \ac{CMQKA}, a novel cross-modal fusion mechanism that achieves linear $O(N)$ complexity through efficient binary operations, enabling scalable hierarchical fusion previously infeasible with conventional attention. \ac{CMQKA} employs bidirectional cross-modal Query-Key attention to extract complementary spatiotemporal features between audio and visual modalities. It uses learnable residual fusion to preserve modality-specific characteristics while enriching representations with cross-modal information. Building upon \ac{CMQKA}, we present SNNergy, an energy-efficient multimodal fusion framework with a three-stage hierarchical architecture that processes audio-visual inputs through progressively decreasing spatial resolutions and increasing semantic abstraction. This multi-scale fusion capability, enabled by \ac{CMQKA}'s linear complexity, allows the framework to capture both local patterns and global context across modalities. Implemented with event-driven binary spike operations, SNNergy achieves remarkable energy efficiency while maintaining fusion effectiveness. Extensive experiments demonstrate that SNNergy establishes new state-of-the-art results on challenging audio-visual benchmarks, including CREMA-D, AVE, and UrbanSound8K-AV, significantly outperforming existing multimodal fusion baselines in both recognition accuracy and computational efficiency. Our  framework\footnote{\url{https://osf.io/7bk9n/overview?view_only=669e7a88cbb74439a6ceb67a22da4280}} advances multimodal fusion by introducing a scalable fusion mechanism that enables hierarchical cross-modal integration with practical energy efficiency for real-world audio-visual intelligence systems.
    \end{abstract}


\end{frontmatter}

\section{Introduction}

Multimodal learning has emerged as a fundamental research direction in artificial intelligence to process and integrate information from multiple sensory modalities such as vision, audio, and text~\cite{baltrusaitis_2019_multimodal_machine_learning, ramachandram_2017_deep_multimodal_learning}. By replicating the human ability to integrate complementary information across modalities, multimodal systems create robust representations that surpass unimodal approaches~\cite{yuille_2006_vision_as_bayesian}. Audio-visual learning, in particular, has demonstrated significant benefits across diverse tasks including speech recognition~\cite{afouras_2018_deep_audio_visual}, emotion recognition~\cite{barros_2016_developing_crossmodal_expression}, event localization~\cite{tian_2018_audio_visual_event_localization}, and action recognition~\cite{kazakos_2019_epic_fusion_audio}. However, effective multimodal fusion faces fundamental challenges, including inherent differences in feature distributions across modalities, varying information quality, and the potential for modality conflicts during integration~\cite{wang_2020_what_makes_training, pham_2019_found_in_translation}.

The success of attention mechanisms in capturing long-range dependencies has naturally led to their adoption for cross-modal fusion~\cite{long_2022_multimodal_attentive_fusion, moorthy_2025_hybrid_multi_attention_network}. Cross-modal attention enables models to selectively focus on relevant features from one modality while processing another, facilitating deep feature interaction and learning of complex cross-modal correlations. However, conventional attention mechanisms suffer from a critical limitation: quadratic computational and memory complexity with respect to the number of tokens. Standard self-attention as used in transformers~\cite{vaswani_2017_attention_is_all} requires $O(N^2)$ operations, where $N$ represents the sequence length. This quadratic scaling creates a fundamental trade-off between fusion effectiveness and computational scalability, severely limiting the construction of hierarchical multi-scale architectures, essential for capturing both local and global patterns in multimodal data~\cite{zhou_2022_spikformer_when_spiking_neural}.

Recent studies have explored linear-complexity alternatives to address this scalability barrier. Query-Key attention mechanisms have demonstrated the possibility of achieving linear $O(N)$ complexity through simplified attention computations~\cite{zhou_2024_qkformer_hierarchical_spiking}. However, these approaches have been primarily developed for unimodal settings and do not directly address the challenges of cross-modal fusion. In contrast, multimodal fusion methods that employ cross-modal attention, such as \ac{SCMRL}~\cite{he_2025_enhancing_audio_visual}, rely on conventional quadratic-complexity attention, which prevents the construction of deep hierarchical architectures with multiple fusion stages at varying spatial resolutions. 
This computational limitation has profound implications for multimodal fusion systems, as hierarchical multi-scale processing has proven essential for robust visual recognition~\cite{he_2016_deep_residual_learning} and has recently been extended to efficient unimodal architectures~\cite{zhou_2024_qkformer_hierarchical_spiking}. Such architectures progressively decrease spatial resolution while increasing semantic abstraction across network stages. However, extending this architectural principle to multimodal fusion requires cross-modal attention at each hierarchical stage, which multiplies computational cost and renders existing approaches impractical for real-world deployment. As a result, the field lacks a cross-modal fusion mechanism that combines the effectiveness of attention-based fusion with the linear computational complexity required for hierarchical multi-scale processing.

To address these challenges, we introduce \ac{CMQKA}, a novel cross-modal fusion mechanism that achieves linear $O(N)$ complexity while enabling effective extraction of complementary spatiotemporal features across modalities. \ac{CMQKA} extends the Query-Key attention paradigm to the cross-modal domain, employing bidirectional attention where queries from one modality attend to keys from another modality using efficient binary operations. Through learnable residual fusion, \ac{CMQKA} preserves modality-specific characteristics while enriching representations with complementary cross-modal information. The linear complexity of \ac{CMQKA} enables the construction of hierarchical multimodal architectures with multiple fusion stages, each operating at progressively coarser spatial scales and higher semantic levels. We implement this mechanism within SNNergy, an energy-efficient multimodal fusion framework that leverages event-driven binary spike operations to further reduce computational overhead while maintaining fusion effectiveness. Our framework processes audio-visual inputs through a three-stage hierarchical architecture, achieving both superior recognition performance and remarkable efficiency compared to existing fusion approaches. Our main contributions are as follows.
\begin{itemize}[nosep,leftmargin=*]
    \item We propose \ac{CMQKA}, a novel cross-modal fusion mechanism that achieves linear $O(N)$ computational complexity while effectively extracting complementary spatiotemporal features across audio and visual modalities. Through bidirectional Query-Key attention and learnable residual fusion, \ac{CMQKA} enables deep cross-modal feature interaction while preserving modality-specific characteristics, addressing the scalability limitations of conventional quadratic-complexity cross-modal attention.

    \item We present SNNergy, an energy-efficient hierarchical multimodal fusion framework that leverages \ac{CMQKA} to enable multi-scale cross-modal integration across three hierarchical stages with progressively decreasing spatial resolutions and increasing semantic abstraction. By implementing fusion through event-driven binary operations, SNNergy achieves practical computational efficiency for real-world audio-visual intelligence systems.

    \item We demonstrate through extensive experiments on audio-visual benchmarks (i.e., UrbanSound8K-AV, AVE, CREMA-D) that our fusion approach establishes new state-of-the-art results, outperforming existing multimodal fusion methods in both classification accuracy and computational efficiency, validating the effectiveness of linear-complexity hierarchical cross-modal attention for scalable multimodal learning.
\end{itemize}

The remainder of this paper is organized as follows. Section~2 reviews related work. Section~3 presents the SNNergy architecture in detail. Section~4 describes the experimental setup, datasets, and implementation details, followed by Section~5 that presents comprehensive results and ablation studies. Section~6 concludes the paper with a summary of our contributions and the broader impact of the proposed framework.


\section{Related Work}

\textbf{Multimodal Fusion and Integration:}
Multimodal learning, particularly audio-visual integration, is inspired by human perception, which naturally combines vision and auditory cues to form coherent interpretations of the environment. By leveraging complementary information across multiple modalities, they have consistently demonstrated advantages over unimodal approaches  \cite{baltrusaitis_2019_multimodal_machine_learning}. 
A central challenge in multimodal learning is therefore how to effectively integrate heterogeneous modalities while capturing cross-modal dependencies and managing computational and energy costs. Early fusion strategies combine features from different modalities at the input level, enabling joint representation learning from the outset (e.g., \cite{ramachandram_2017_deep_multimodal_learning}). Late fusion approaches process each modality independently and combine their predictions at the decision level (e.g., \cite{gadzicki_2020_early_vs_late}). Intermediate fusion integrates modalities at multiple stages of the network, balancing representational depth and flexibility (e.g., \cite{liu_2024_multimodal_alignment_and}). While these classical approaches offer different trade-offs between integration depth and computational overhead, they typically rely on simple concatenation or weighted averaging, limiting their ability to model the complex inter-modal dependencies present in audio-visual data.

Attention-based fusion mechanisms address this limitation by dynamically weighting modality contributions and explicitly modeling cross-modal correspondence \cite{long_2022_multimodal_attentive_fusion, moorthy_2025_hybrid_multi_attention_network}. Cross-modal attention allows features from one modality to guide representation learning in another, facilitating precise alignment of audio and visual events \cite{arandjelovic_2018_objects_that_sound}. Relatedly, cross-modal distillation transfers knowledge from a strong modality to a weaker one to improve robustness \cite{gupta_2022_cross_modal_distillation}. However, standard scaled dot-product attention inherits quadratic computational complexity, making it prohibitively expensive for high-resolution or resource-constrained multimodal inputs.

Beyond computational complexity, audio-visual fusion presents additional challenges. Temporal alignment is nontrivial due to differing temporal resolutions and sampling rates \cite{arandjelovic_2017_look_listen_and}. Modality imbalance, where one modality dominates learning, can degrade fusion quality if not carefully addressed \cite{wang_2020_what_makes_training}. Real-world noise and signal variability further complicate robust multimodal recognition \cite{hu_2018_squeeze_and_excitation}. Crucially, the energy consumption of deep multimodal fusion models limits their applicability on edge and mobile platforms, motivating the development of energy-efficient fusion architectures that preserve accuracy while reducing computational overhead.

\textbf{Energy-Efficient Multimodal Fusion:}
Addressing energy efficiency in multimodal fusion requires rethinking both the computational substrate and fusion mechanism. Event-driven computing paradigms, particularly \acp{SNN}, offer a promising direction by replacing dense floating-point operations with sparse binary spike events that trigger computation only when necessary \cite{maass_1997_networks_of_spiking}. This event-driven nature enables dramatic energy savings: neuromorphic hardware implementations of \acp{SNN} can achieve orders-of-magnitude reduction in power consumption compared to conventional GPU-based inference \cite{davies_2018_loihi_a_neuromorphic}. However, translating effective multimodal fusion mechanisms into the spiking domain while preserving both fusion quality and computational efficiency remains an open challenge. 

Early audio-visual \acp{SNN} fusion methods relied on simple concatenation strategies \cite{kim_2021_audio_visual_spiking}, which proved insufficient for capturing cross-modal complementarity. 
More recent work has explored cross-modal attention mechanisms in energy-efficient settings. The \ac{SCMRL} framework \cite{he_2025_enhancing_audio_visual} introduces the \ac{CCSSA} mechanism to extract complementary audio-visual features through spatiotemporal cross-modal attention. \ac{CCSSA} decomposes cross-modal fusion into spatial (\ac{SCSA}) and temporal (\ac{TCSA}) components, integrating complementary features via residual connections while preserving modality-specific discriminative power. \ac{SCMRL} further employs contrastive alignment to reduce distributional shifts between modalities, achieving strong performance on audio-visual benchmarks (CREMA-D, UrbanSound8K-AV) while leveraging event-driven spiking computation for energy efficiency \cite{he_2025_enhancing_audio_visual}. 
Despite these advances, SCMRL relies on conventional spiking self-attention with quadratic computational complexity, where computational cost scales with the square of the number of spatial tokens. This fundamentally prevents the construction of hierarchical multi-scale fusion architectures, as cross-modal attention at multiple resolutions becomes computationally prohibitive. Consequently, SCMRL is restricted to single-scale non-hierarchical fusion, limiting its ability to capture audio-visual correspondences across multiple spatial and temporal scales—a capability shown to be essential in state-of-the-art vision architectures \cite{liu_2021_swin_transformer_hierarchical}. This exposes a key gap in current energy-efficient multimodal fusion approaches: the absence of a linear -complexity cross-modal attention mechanism capable of supporting scalable hierarchical fusion.

Complementary efforts have explored multimodal \ac{SNN} architectures through innovations in neuron-level fusion and temporal attention mechanisms. MISNet \cite{liu_2025_towards_energy_efficient} introduces the Multimodal Leaky Integrate-and-Fire neuron, which coordinates spike activations across audio and visual modalities within a single neuron, addressing cross-modal asynchrony. 
Through multi-round spike interaction and modality-specific loss regularization, MISNet demonstrates improved balance between accuracy and energy efficiency on audio-visual classification tasks. Complementary efforts on this neuron-level innovation proposed temporal attention-guided adaptive fusion \cite{shen_2025_spiking_neural_networks}, which addresses modality imbalance by dynamically assigning importance scores to fused features at each timestep. This approach enables hierarchical integration of temporally heterogeneous spike-based features while modulating learning rates per modality based on attention scores, preventing dominant modalities from monopolizing optimization. These studies highlight the importance of both neuron-level mechanisms and temporal dynamics, but do not address the scalability limitations imposed by quadratic attention. 

\textbf{Spiking Transformers and Hierarchical Learning:}
While the above methods focus on multimodal fusion, recent advances in unimodal spiking transformers provide crucial insights into scalable and hierarchical attention mechanisms in the spiking domain. Spikformer~\cite{zhou_2022_spikformer_when_spiking_neural} introduced \ac{SSA}, which models sparse visual features using spike-form Query, Key, and Value representations without softmax operations. By leveraging the binary nature of spikes, \ac{SSA} replaces costly multiplication operations with simple masking and accumulation, achieving substantial energy savings while maintaining competitive performance on both static and neuromorphic datasets. 
Subsequent work further reduced non-spiking computation and improved efficiency through spike-driven residual connections \cite{zhou_2023_spikingformer_spike_driven} and masking-based addition operations that eliminate multiplication entirely \cite{yao_2023_spike_driven_transformer}. 

However, \ac{SSA} retains quadratic complexity, limiting its applicability to hierarchical architectures with multi-scale representations. QKFormer \cite{zhou_2024_qkformer_hierarchical_spiking} addresses this limitation by introducing Query-Key attention with linear complexity. Unlike \ac{SSA}, which requires Query, Key, and Value components, Q-K attention uses only Query and Key to compute attention weights through binary spike vectors. 
By aggregating attention through row- and column-wise operations (Q-K Token Attention and Q-K Channel Attention), QKFormer enables efficient multi-scale representation learning. Together with the \ac{SPEDS} module to improve spiking information transmission across downsampling blocks, QKFormer achieved state-of-the-art performance (exceeding 85\% top-1 accuracy on ImageNet-1K) and was the first directly trained \ac{SNN}  \cite{zhou_2024_qkformer_hierarchical_spiking}.

\textbf{Positioning of Our Work:}
To bridge the gap in scalable, energy-efficient multimodal fusion, we propose SNNergy. This hierarchical audio-visual fusion framework introduces \ac{CMQKA}, the first cross-modal fusion mechanism achieving linear $O(N)$ complexity while preserving effective complementary feature extraction. SNNergy integrates insights from QKFormer, which demonstrates that linear-complexity Query-Key attention enables hierarchical multi-scale architectures infeasible with quadratic attention \cite{zhou_2024_qkformer_hierarchical_spiking}, while SCMRL establishes that residual integration of cross-modal complementary features preserves modality-specific discriminative power \cite{he_2025_enhancing_audio_visual}. By extending linear Query-Key attention to the cross-modal domain, SNNergy enables scalable hierarchical multimodal fusion previously infeasible in energy-efficient settings. 

\section{Preliminary Definitions}
In this section, we introduce the problem definition, a brief overview of the main components of our SNNergy framework, including the \ac{LIF} neuron model, and the Query-Key attention mechanism adapted to multimodality.
\subsection{Problem Definition}
Consider a multimodal learning task where the objective is to learn a mapping from paired audio-visual inputs to a set of semantic labels. Let $\mathcal{D} = \{(\mathbf{X}_i^a, \mathbf{X}_i^v, y_i)\}_{i=1}^{N}$ denote the training dataset consisting of $N$ samples, where $\mathbf{X}_i^a \in \mathbb{R}^{T_a \times F_a}$ represents the audio input with $T_a$ temporal frames and $F_a$ frequency bins (e.g., mel-spectrogram features), $\mathbf{X}_i^v \in \mathbb{R}^{T_v \times H \times W \times 3}$ represents the visual input comprising $T_v$ frames of height $H$ and width $W$ with RGB channels, and $y_i \in \{1, 2, \ldots, C\}$ denotes the ground-truth label from $C$ semantic categories.
The primary objective is to learn a multimodal function $f_{\theta}: \mathcal{X}^a \times \mathcal{X}^v \rightarrow \mathcal{Y}$ parameterized by $\theta$, which maps the joint audio-visual input space $\mathcal{X}^a \times \mathcal{X}^v$ to the label space $\mathcal{Y} = \{1, \ldots, C\}$. Formally, the learning problem can be expressed as:
\begin{equation}
    \theta^* = \arg\min_{\theta} \sum_{i=1}^{N} \mathcal{L}(f_{\theta}(\mathbf{X}_i^a, \mathbf{X}_i^v), y_i) + \lambda \Omega(\theta),
\end{equation}
where $\mathcal{L}(\cdot, \cdot)$ is a loss function (e.g., cross-entropy loss for classification), $\Omega(\theta)$ represents a regularization term, and $\lambda$ controls the regularization strength.

This paper focuses on designing the multimodal function $f_{\theta}$ using \ac{SNN} architectures that effectively integrate audio and visual information while leveraging the temporal dynamics inherent in spike-based processing. The proposed SNNergy framework aims to address the challenges of cross-modal fusion and efficient representation learning within the spiking domain.

\subsection{\ac{LIF} Neuron Model}
The \ac{LIF} neuron serves as a fundamental computational unit in \acp{SNN}, providing biologically-plausible temporal dynamics for spike-based information processing. At each time step $t$, the membrane potential $V_i^t$ of the $i^{th}$ neuron evolves according to the discrete-time leaky integration dynamics:
\begin{equation}
    V_i^t = \tau V_i^{t-1}(1 - S_i^{t-1}) + W_i X_i^t, \label{dynamicsEq}
\end{equation}
where $\tau \in (0, 1)$ is the membrane decay constant controlling the leakage rate, $S_i^{t-1} \in \{0, 1\}$ is the binary spike output from the previous time step, $W_i$ represents the synaptic weights, and $X_i^t$ denotes the input current at time $t$. The neuron generates a spike when the membrane potential exceeds a threshold $V_{\text{th}}$, implemented through the Heaviside step function:
\begin{equation}
    S_i^t = H(V_i^t - V_{\text{th}}) = \begin{cases}
        1, & \text{if } V_i^t \geq V_{\text{th}}, \\
        0, & \text{otherwise}.
    \end{cases}
\end{equation}
Following spike emission, the membrane potential undergoes a hard reset as indicated by the $(1 - S_i^{t-1})$ term in equation~\ref{dynamicsEq}. During backpropagation, the non-differentiable Heaviside function is approximated using a surrogate gradient function to enable gradient-based training~\cite{neftci_2019_surrogate_gradient_learning, wu_2018_spatio_temporal_backpropagation}.

\subsection{Self Attention in Spiking Neural Networks}

The transformer architecture relies fundamentally on the self-attention mechanism to capture long-range dependencies and model complex relationships within input sequences. In the vanilla self-attention mechanism~\cite{vaswani_2017_attention_is_all}, three floating-point components, Query ($\mathbf{Q}_F$), Key ($\mathbf{K}_F$), and Value ($\mathbf{V}_F$), are derived from the input through learnable linear projections: $\mathbf{Q}_F, \mathbf{K}_F, \mathbf{V}_F = \mathbf{X}(\mathbf{W}_Q, \mathbf{W}_K, \mathbf{W}_V)$. The attention output is computed as:
\begin{equation}
    \text{VSA}(\mathbf{Q}_F, \mathbf{K}_F, \mathbf{V}_F) = \text{Softmax}\left(\frac{\mathbf{Q}_F\mathbf{K}_F^T}{\sqrt{d}}\right)\mathbf{V}_F,
\end{equation}
where $d$ is the dimension of the key vectors and the scaling factor $1/\sqrt{d}$ normalizes the attention scores. However, this formulation presents significant challenges for \acp{SNN}: the softmax operation involves exponential calculations and division, while the matrix multiplications operate on continuous floating-point values, both of which are incompatible with the discrete, event-driven nature of spike-based computation.

To address these incompatibilities, \ac{SSA} reformulates the attention mechanism to operate entirely with binary spike representations~\cite{zhou_2022_spikformer_when_spiking_neural}. The spike-form Query, Key, and Value matrices are generated through learnable linear layers followed by batch normalization and spiking neurons: $\mathbf{I} = \text{SN}_I(\text{BN}_I(\mathbf{X}\mathbf{W}_I))$ for $\mathbf{I} \in \{\mathbf{Q}, \mathbf{K}, \mathbf{V}\}$, where $\mathbf{Q}, \mathbf{K}, \mathbf{V} \in \{0, 1\}^{T \times N \times D}$. The \ac{SSA} computation is then formulated as:
\begin{equation}
    \text{SSA}(\mathbf{Q}, \mathbf{K}, \mathbf{V}) = \text{SN}\left(\mathbf{Q}\mathbf{K}^T\mathbf{V} \times s\right),
\end{equation}
where $s$ is a learnable scaling factor, and SN denotes the spiking neuron layer. By replacing continuous multiplications with sparse binary operations and eliminating the softmax function, \ac{SSA} achieves substantial energy efficiency while preserving the core attention mechanism. However, the computational complexity of \ac{SSA} remains $O(N^2D)$, scaling quadratically with the number of tokens $N$, which severely constrains the construction of hierarchical architectures with multi-scale feature representations.

\subsection{Query-Key Linear Attention}

To overcome the quadratic complexity bottleneck of \ac{SSA}, we adopt the Query-Key attention mechanism~\cite{zhou_2024_qkformer_hierarchical_spiking} to achieve linear computation while operating exclusively with binary spike operations. Given spike-form input $\mathbf{X} \in \{0, 1\}^{T \times N \times D}$ with $N$ tokens and $D$ channels, the Query and Key matrices, $\mathbf{Q}, \mathbf{K} \in \{0, 1\}^{T \times N \times D}$, are generated through learnable linear transformations followed by batch normalization and \ac{LIF} neurons: $\mathbf{Q} = \text{SN}_Q(\text{BN}(\mathbf{X}\mathbf{W}_Q))$ and $\mathbf{K} = \text{SN}_K(\text{BN}(\mathbf{X}\mathbf{W}_K))$. In the \ac{QKTA} variant, a token attention vector $\mathbf{A}_t \in \{0, 1\}^{N \times 1}$ is computed by row-wise summation over the channel dimension followed by a spiking neuron $\mathbf{A}_t = \text{SN}(\sum_{i=0}^{D} \mathbf{Q}_{i,j})$, which models the binary importance of different tokens. The output is then obtained through a column-wise mask operation (Hadamard product): $\mathbf{X}' = \mathbf{A}_t \otimes \mathbf{K}$, followed by a post-linear layer $\mathbf{X}'' = \text{SN}(\text{BN}(\text{Linear}(\mathbf{X}')))$. This mechanism achieves $O(N)$ or $O(D)$ complexity depending on the attention dimension~\cite{zhou_2024_qkformer_hierarchical_spiking}, enabling hierarchical architectures with progressively decreasing spatial resolution across stages. 

\section{SNNergy Framework}
\subsection{Cross-Modal Query-Key Attention}
As mentioned earlier, existing cross-modal attention mechanisms  in multimodal spiking neural networks operate on the $O(N^2)$ complexity paradigm~\cite{he_2025_enhancing_audio_visual}, restricting their use to single-scale, hierarchical designs. To overcome this limitation, we introduce the \ac{CMQKA} module, which extends the linear-complexity Q-K attention mechanism to the multimodal domain.
Our key insight is that cross-modal complementarity can be captured through Query-Key projections: one modality provides the Query that encodes ``what information to seek,'' while the other modality provides the Key representing ``what information is available.'' By decomposing attention into separate spatial and temporal attentions, and leveraging binary spike operations throughout, \ac{CMQKA} achieves $O(N)$ complexity while extracting rich complementary spatiotemporal features between audio and visual modalities. These cross-modal features are then integrated with the original modality-specific representations through learnable residual connections, preserving modality-specific power while enriching representations with complementary information. For clarity, we present the mechanism for a single sample with single-head attention; multi-head extensions follow standard practice.

\begin{figure}[t]
    \centering
    \includegraphics[width=\textwidth]{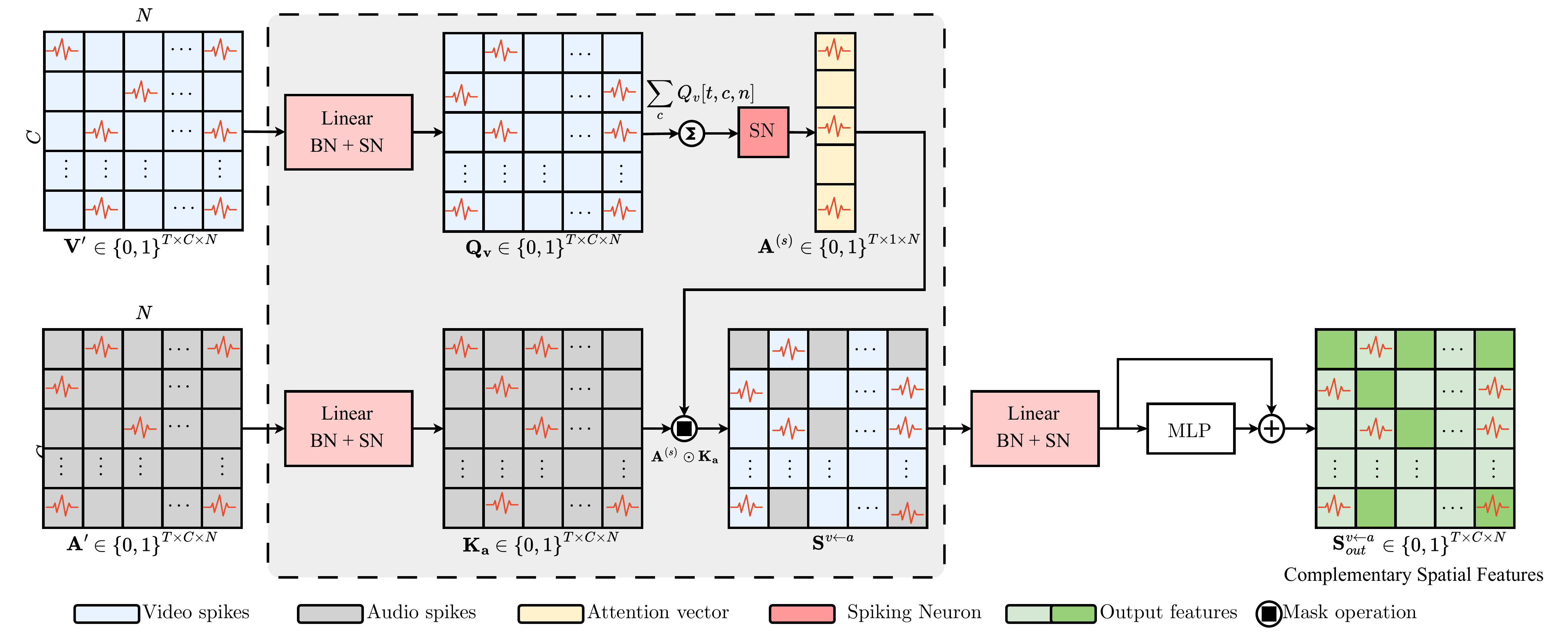}
    \caption{Illustration of the Complementary Spatial Attention pathway within the \ac{CMQKA} module: Video spikes $\mathbf{V}' \in \{0,1\}^{T \times B \times C \times N}$ and audio spikes $\mathbf{A}' \in \{0,1\}^{T \times B \times C \times N}$ are transformed through learnable projections to generate Query $\mathbf{Q}_v$ and Key $\mathbf{K}_a$, both in $\{0,1\}^{T \times B \times C \times N}$. For multi-head attention, Q is reshaped to $\{0,1\}^{T \times B \times h \times C_h \times N}$ where $h$ is the number of heads and $C_h = C/h$ is the per-head channel dimension. The Query is summed over the per-head channel dimension $C_h$ and passed through a spiking neuron (SN) to produce a binary spatial attention tensor $\mathbf{A}^{(s)} \in \{0,1\}^{T \times B \times h \times 1 \times N}$. This attention tensor selectively masks the Key matrix through element-wise multiplication ($\mathbf{S}^{v \leftarrow a} = \mathbf{A}_t^{(s)} \odot \mathbf{K}_a$) to extract complementary spatial features from the audio modality that are relevant to the video modality. The mechanism maintains $O(N)$ linear complexity through binary spike operations, avoiding the quadratic cost of traditional self-attention.}
    \label{fig:complementary_spatial_attention}
\end{figure}

\paragraph{\textbf{Complementary Spatial Attention}}
Spatial correspondence between audio and visual modalities is fundamental to audio-visual scene understanding. However, determining which spatial regions in one modality complement features in another requires selective attention mechanisms. As illustrated in \Cref{fig:complementary_spatial_attention}, our Complementary Spatial Attention pathway addresses this challenge by operating on the token dimension to identify and extract spatially-aligned cross-modal features.

Given spike-form video and audio inputs $\mathbf{V}, \mathbf{A} \in \{0, 1\}^{T \times C \times H \times W}$, we flatten the spatial dimensions to obtain $\mathbf{V}', \mathbf{A}' \in \{0, 1\}^{T \times C \times N}$ with $N = H \times W$ tokens. To enable cross-modal spatial attention, we employ asymmetric projections: the Query is derived from the video modality to encode spatial patterns of interest, while the Key is derived from the audio modality to represent available auditory information. Both are generated through modality-specific learnable transformations:
\begin{equation}
    \mathbf{Q}_v = \text{SN}_Q(\text{BN}(\text{Conv1D}(\mathbf{V}', \mathbf{W}_Q^v))), \quad \mathbf{K}_a = \text{SN}_K(\text{BN}(\text{Conv1D}(\mathbf{A}', \mathbf{W}_K^a))),
\end{equation}
where $\mathbf{Q}_v, \mathbf{K}_a \in \{0, 1\}^{T \times C \times N}$. This asymmetric design enables the network to learn which audio tokens are spatially relevant to specific visual content.
To achieve linear complexity, we compute a binary spatial attention mask by aggregating the Query matrix over channels: 
\begin{equation}
    \mathbf{A}_t^{(s)} = \text{SN}\left(\sum_{c=1}^{C} \mathbf{Q}_{v}[t, c, :]\right), \quad \mathbf{A}_t^{(s)} \in \{0, 1\}^{T \times 1 \times N},
\end{equation}
where the channel-wise summation captures the overall importance of each spatial token, and the spiking neuron converts accumulated membrane potentials into binary attention weights. This attention vector identifies which audio tokens are complementary to the video modality at each timestep. We then extract complementary spatial features through selective masking:
\begin{equation}
    \mathbf{S}^{v \leftarrow a} = \mathbf{A}_t^{(s)} \odot \mathbf{K}_a,
\end{equation}
where the Hadamard product $\odot$ (with broadcasting) retains only those audio tokens deemed relevant by the video-derived attention mask, yielding $\mathbf{S}^{v \leftarrow a} \in \{0, 1\}^{T \times C \times N}$, representing the complementary spatial features extracted from the audio modality.

\begin{figure}[t]
    \centering
    \includegraphics[width=\textwidth]{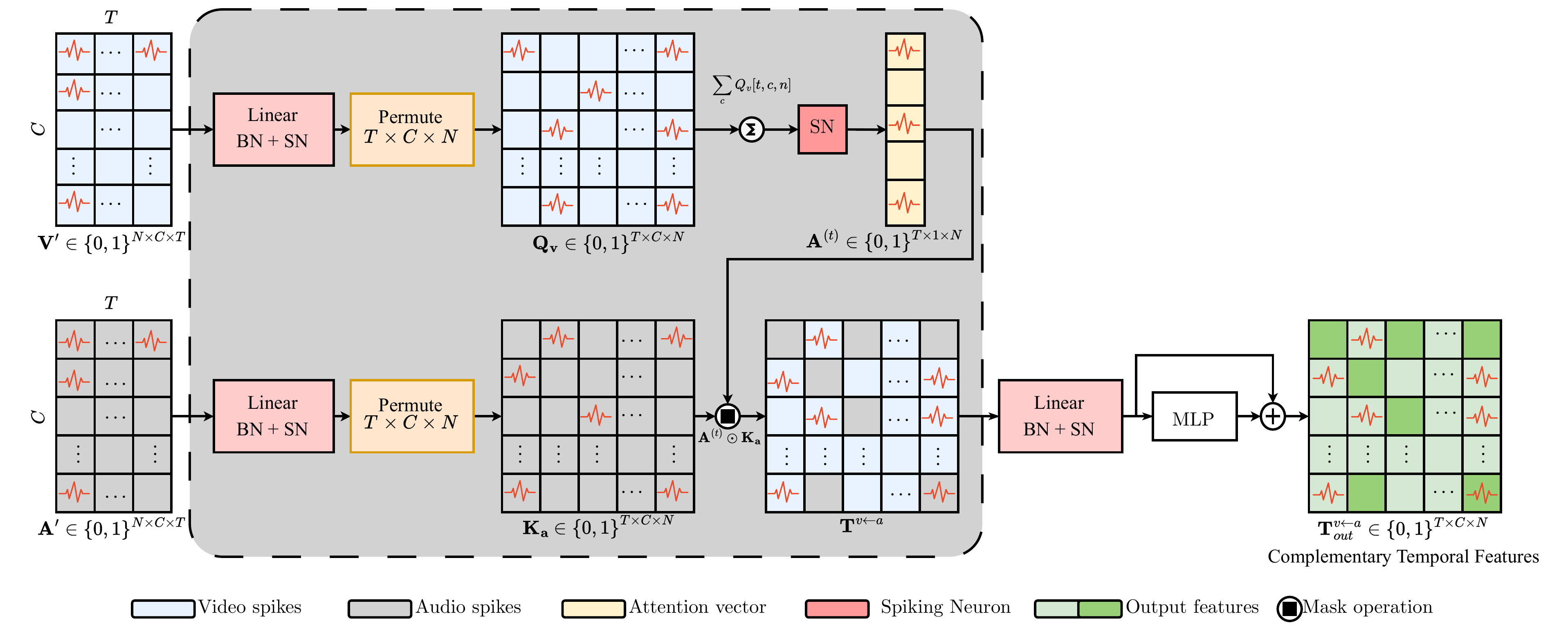}
    \caption{Illustration of the Complementary Temporal Attention pathway within the \ac{CMQKA} module: The spatially-flattened inputs $\mathbf{V}', \mathbf{A}' \in \{0,1\}^{T \times C \times N}$ are first permuted to $N \times C \times T$ to operate along the temporal dimension. The permuted inputs are transformed through learnable projections to generate Query $\mathbf{Q}_v$ from video and Key $\mathbf{K}_a$ from audio, both in $\{0,1\}^{N \times C \times T}$ and then permuted back to $T \times C \times N$. The Query matrix is summed over channels and passed through a spiking neuron to produce a binary temporal attention vector $\mathbf{A}^{(t)} \in \{0,1\}^{T \times 1 \times N}$. This attention vector selectively masks the Key matrix through element-wise multiplication ($\mathbf{T}^{v \leftarrow a} = \mathbf{A}_t^{(t)} \odot \mathbf{K}_a^{(t)}$) to extract complementary temporal features $\mathbf{T}^{v \leftarrow a}$ from the audio modality that capture temporal patterns relevant to the video modality. The mechanism maintains $O(N)$ linear complexity, enabling efficient temporal cross-modal feature extraction.}
    \label{fig:complementary_temporal_attention}
\end{figure}

\paragraph{\textbf{Complementary Temporal Attention}}
While spatial attention captures where cross-modal interactions occur, temporal dynamics are equally critical in audio-visual learning. Audio and visual events unfold over time with complex synchronization patterns (e.g., speech articulation aligns with lip movements, or sound onsets coincide with visual actions). Our Complementary Temporal Attention pathway exploits these temporal dependencies by identifying features in one modality that provide complementary information to the other (\Cref{fig:complementary_temporal_attention}).

To extract temporal complementarity, we operate along the time dimension by permuting the flattened inputs $\mathbf{V}', \mathbf{A}' \in \{0, 1\}^{T \times C \times N}$ to $N \times C \times T$, treating time as a sequence dimension. We then compute asymmetric temporal projections:
\begin{align}
\begin{split}
    \mathbf{Q}_v^{(t)} &= \text{SN}_Q(\text{BN}(\text{Conv1D}(\mathbf{V}'_{\text{perm}}, \mathbf{W}_Q^{v,t}))), \\ \mathbf{K}_a^{(t)} &= \text{SN}_K(\text{BN}(\text{Conv1D}(\mathbf{A}'_{\text{perm}}, \mathbf{W}_K^{a,t}))),
\end{split}
\end{align}
where $\mathbf{Q}_v^{(t)}, \mathbf{K}_a^{(t)} \in \{0, 1\}^{N \times C \times T}$ are then permuted back to $T \times C \times N$ for consistency. This formulation enables the network to learn which temporal patterns in the audio stream are relevant to the temporal evolution of visual content.
As shown in \Cref{fig:complementary_temporal_attention}, we compute the temporal attention mask through channel aggregation, analogous to the spatial pathway:
\begin{equation}
    \mathbf{A}_t^{(t)} = \text{SN}\left(\sum_{c=1}^{C} \mathbf{Q}_{v}^{(t)}[t, c, :]\right), \quad \mathbf{A}_t^{(t)} \in \{0, 1\}^{T \times 1 \times N},
\end{equation}
yielding per-timestep binary masks that identify complementary temporal patterns between modalities. The complementary temporal features are then extracted via selective masking:
\begin{equation}
    \mathbf{T}^{v \leftarrow a} = \mathbf{A}_t^{(t)} \odot \mathbf{K}_a^{(t)},
\end{equation}
where $\mathbf{T}^{v \leftarrow a} \in \{0, 1\}^{T \times C \times N}$ captures temporal dependencies between audio and visual streams, representing the complementary temporal features extracted from the audio modality.

\begin{figure}[t]
    \centering
    \includegraphics[width=0.92\textwidth]{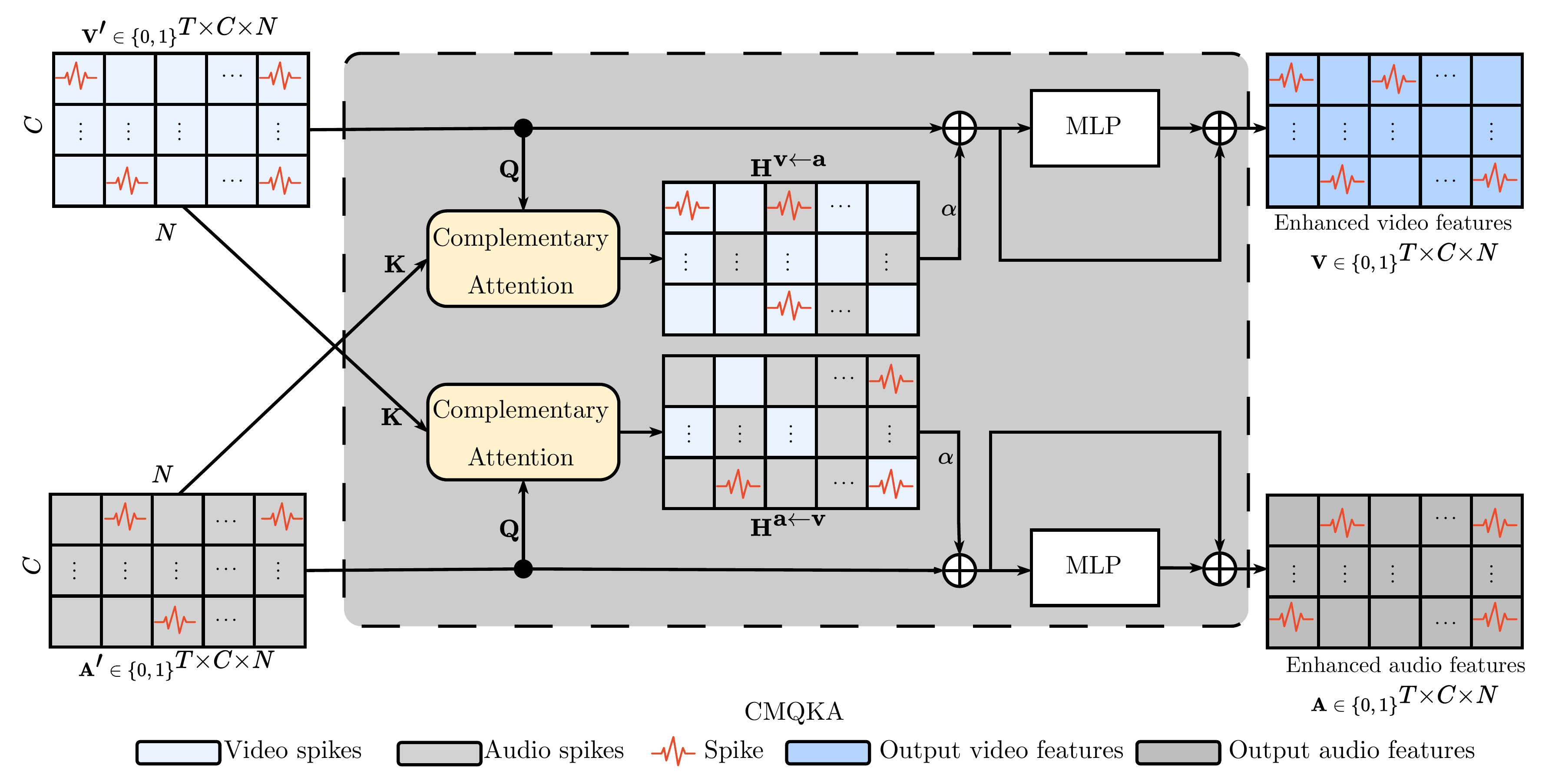}
    \caption{Fusion and residual integration within the \ac{CMQKA} module: The complementary spatial features $\mathbf{S}^{v \leftarrow a}$ and temporal features $\mathbf{T}^{v \leftarrow a}$ are pooled along their respective dimensions to produce reduced representations $\mathbf{S}_{\text{red}}^{v \leftarrow a}$ and $\mathbf{T}_{\text{red}}^{v \leftarrow a}$. Element-wise multiplication combines these reduced features into joint spatiotemporal complementary features $\mathbf{H}^{v \leftarrow a}$. A learnable residual connection integrates these cross-modal features into the original video representation $\mathbf{V}$, yielding enriched output features $\tilde{\mathbf{V}}$. A spiking MLP with residual connection further refines the fused features to produce the final output $\mathbf{V}_{\text{out}}$. This fusion strategy preserves modality-specific characteristics while enhancing representations with cross-modal complementary information.}
    \label{fig:cmqka_fusion}
\end{figure}

\paragraph{\textbf{Cross-Modal Feature Fusion and Residual Integration}}
Having extracted complementary features along both spatial and temporal dimensions, we address a critical design question: ``how should these two pathways be combined to form a unified cross-modal representation?'' Naive concatenation or summation fails to model the interaction between spatial and temporal complementarity. Instead, we propose a multiplicative fusion strategy that captures the joint spatiotemporal relevance of cross-modal features, as depicted in \Cref{fig:cmqka_fusion}.

We first apply dimension-specific pooling to compress each pathway's output. Spatial complementary features $\mathbf{S}^{v \leftarrow a}$ are averaged over spatial dimensions to yield $\mathbf{S}_{\text{red}}^{v \leftarrow a} \in \{0, 1\}^{T \times C \times 1 \times 1}$, retaining temporal and channel information while abstracting over space. Conversely, temporal complementary features $\mathbf{T}^{v \leftarrow a}$ are averaged over time to obtain $\mathbf{T}_{\text{red}}^{v \leftarrow a} \in \{0, 1\}^{1 \times C \times H \times W}$, preserving spatial structure while aggregating temporal patterns. We then fuse these reduced representations through element-wise multiplication:
\begin{equation}
    \mathbf{H}^{v \leftarrow a} = \mathbf{S}_{\text{red}}^{v \leftarrow a} \odot \mathbf{T}_{\text{red}}^{v \leftarrow a},
\end{equation}
where broadcasting produces $\mathbf{H}^{v \leftarrow a} \in \{0, 1\}^{T \times C \times H \times W}$. This multiplicative fusion ensures that only features exhibiting both spatial and temporal complementarity receive high activation, creating a joint spatiotemporal cross-modal representation.

A critical challenge in multimodal fusion is balancing modality-specific features with cross-modal complementary information. Aggressive fusion can dilute discriminative modality-specific patterns, while insufficient integration fails to leverage cross-modal synergy. Following the residual learning paradigm established by SCMRL~\cite{he_2025_enhancing_audio_visual}, we integrate cross-modal features through a learnable residual connection:
\begin{equation}
    \tilde{\mathbf{V}} = \mathbf{V} + \alpha \cdot \mathbf{H}^{v \leftarrow a},
\end{equation}
where $\alpha$ is a learnable scalar (initialized to 1.5 in our experiments). This formulation preserves the original modality-specific features $\mathbf{V}$ while adding cross-modal complementary information $\mathbf{H}^{v \leftarrow a}$ as a residual term. The learnable parameter $\alpha$ enables the network to adaptively determine the optimal balance between unimodal and cross-modal representations based on task requirements and data characteristics.

To further enhance representational capacity, we apply a final spiking \ac{MLP} with residual connection: $\mathbf{V}_{\text{out}} = \tilde{\mathbf{V}} + \text{MLP}(\tilde{\mathbf{V}})$, enabling nonlinear feature transformation while maintaining spike-form processing throughout.

By operating bidirectionally (i.e., computing both $\mathbf{H}^{v \leftarrow a}$ for the video pathway and $\mathbf{H}^{a \leftarrow v}$ for the audio pathway), the \ac{CMQKA} module ensures symmetric information flow between modalities. Each modality benefits equally from cross-modal complementarity while maintaining linear computational complexity of $O(N)$, a critical property in deep hierarchical architectures construction, infeasible with quadratic-complexity attention mechanisms.

\subsection{Deep Hierarchical Multimodal Spiking Transformer}

While the \ac{CMQKA} module provides efficient cross-modal feature extraction, achieving state-of-the-art performance on complex audio-visual tasks requires the ability to capture features at multiple spatial and temporal scales. Hierarchical architectures have proven essential in vision models~\cite{he_2016_deep_residual_learning, liu_2021_swin_transformer_hierarchical} and unimodal spiking networks~\cite{zhou_2024_qkformer_hierarchical_spiking}, as they enable progressive semantic abstraction through multi-scale feature representations. However, constructing hierarchical multimodal \acp{SNN} has been fundamentally constrained by the quadratic complexity of traditional \ac{SSA}. This quadratic bottleneck makes it computationally prohibitive to process high-resolution inputs across multiple stages with progressively decreasing spatial dimensions.

The linear $O(N)$ complexity of our \ac{CMQKA} mechanism fundamentally eliminates this barrier, enabling, for the first time, the construction of deep hierarchical spiking transformers for multimodal learning. Building upon this foundation, we propose SNNergy, a three-stage hierarchical architecture that processes audio and visual modalities through separate but interacting pathways. Each stage combines spatial downsampling with cross-modal attention, enabling the network to extract features at progressively coarser spatial scales while increasing channel capacity and semantic abstraction. This hierarchical design mirrors the success of multi-scale processing in state-of-the-art vision models while maintaining the energy efficiency and temporal dynamics inherent to spike-based computation.

\paragraph{\textbf{Architecture Overview}}
The SNNergy framework adopts a deep hierarchical architecture designed to process audio-visual data through progressively abstract semantic levels, mirroring the multi-scale processing of advanced vision systems. Unlike conventional flat fusion models, SNNergy operates across three distinct stages, each serving a specific role in the transition from fine-grained detail to global semantic understanding.

The architecture processes video and audio inputs through two parallel, interacting pathways that progressively reduce spatial resolution while expanding channel capacity. Following standard hierarchical design principles, channel dimensions grow proportionally at each stage as spatial resolution decreases, enabling the network to capture increasingly abstract semantic features. This hierarchical design is governed by two primary functional units: the \ac{SPDS} module for efficient downsampling and encoding, and cross-modal attention blocks for feature integration. The architecture stages are as follows: 

\textbf{Stage 1 (High-Resolution Feature Fusion).} The initial stage focuses on extracting and fusing fine-grained spatiotemporal patterns. Raw inputs are first encoded into spike-form representations by modality-specific \ac{SPDS} stems to perform initial spatial downsampling and create patch embeddings suitable for transformer processing. To enable efficient cross-modal interaction at this high-resolution scale, we employ a stack of \ac{CMQKA} blocks. These blocks leverage linear-complexity attention to facilitate dense bidirectional information flow between modalities without incurring the prohibitive computational costs of standard self-attention on large feature maps.

\textbf{Stage 2 (Intermediate Representation Learning).} The second \ac{SPDS} module further downsample the feature maps while doubling the channel depth to abstract the local features into mid-level semantic concepts. A second series of \ac{CMQKA} blocks continues the fusion process, refining the intermediate representations. This stage acts as a crucial bridge, transforming local details into coherent semi-global features while maintaining energy-efficient linear complexity.

\textbf{Stage 3 (Global Semantic Fusion).} The final stage is designed to capture high-level semantic dependencies. At this coarsest spatial resolution, the substantially reduced token count permits the strategic deployment of global attention mechanisms. Consequently, we utilize cross-modal \ac{SSA} blocks rather than linear attention. This hybrid design choice allows SNNergy to capture long-range global contexts that span the entire receptive field, essential for accurate classification, while restricting the computationally expensive quadratic complexity to only the smallest feature maps.

Following the hierarchical processing, the refined modality-specific representations are aggregated via Global Average Pooling (GAP) to produce compact feature vectors. These vectors are fused to yield a unified multimodal representation for final semantic prediction.

\paragraph{\textbf{Spiking Patch Downsampling with Shortcut}}
Unlike continuous-valued networks, where pooling layers retain magnitude information, binary spike maps are susceptible to significant information loss when spatial resolution is reduced. To mitigate this and ensure robust feature propagation across hierarchical levels, we employ the \ac{SPDS} module. This module serves as the primary encoding and downsampling unit, designed to abstract spatial dimensions while preserving the integrity of the temporal spike train.

The \ac{SPDS} architecture is structured as a residual block comprising two parallel pathways: a deep feature extraction branch and a lightweight shortcut connection. The extraction branch utilizes a sequence of convolutional layers interleaved with Batch Normalization (BN) and Max-Pooling operations to capture local dependencies and progressively reduce spatial resolution. Crucially, \ac{LIF} neurons are embedded within this pathway to ensure that the feature transformation remains strictly within the event-driven domain, converting analog convolutional outputs into sparse binary representations.

Simultaneously, the shortcut pathway acts as a preservation mechanism. It applies an aligned strided transformation, matching the spatial reduction and channel expansion of the main branch to facilitate efficient gradient propagation during backpropagation. This directly addresses the vanishing gradient problem often encountered in deep spiking architectures. The outputs of these two branches are aggregated via element-wise addition before passing through a final spiking activation function. Formally, for an input feature map $\mathbf{X}_{in}$, the module generates the down sampled output $\mathbf{X}_{out}$ as:
\begin{equation}
    \mathbf{X}_{out} = \text{LIF}\left( \mathcal{F}_{ext}(\mathbf{X}_{in}) + \mathcal{F}_{skip}(\mathbf{X}_{in}) \right),
\end{equation}
where $\mathcal{F}_{ext}$ and $\mathcal{F}_{skip}$ denote the deep extraction operations and the residual projection. By integrating information from both the semantic extraction branch and the structural shortcut, \ac{SPDS} ensures that subsequent attention blocks receive high-quality, information-dense spike representations, regardless of the reduction factor (e.g., $4\times$ in the stem or $2\times$ in intermediate stages).

\section{Experimental Design and Analysis}
\label{sec:experiment}
\subsection{Experimental Setup}
\paragraph{\textbf{Benchmark datasets}} We test our framework's performance on three challenging audio-visual public benchmarks: 
\begin{itemize}[leftmargin=*,nosep]
    \item The AVE (Audio-Visual Event) dataset~\cite{tian_2018_audio_visual_event_localization} consists of 4,143 videos covering 28 event categories, including human activities, animal sounds, musical instruments, and environmental sounds. While originally designed for temporal event localization, we adapt AVE for audio-visual event classification, where the goal is to predict the event category from the multimodal input. The official split allocates 3,339 videos for training, with the remaining samples equally divided into validation and test sets. All inputs use $128 \times 128$ spatial resolution for both visual frames and mel-spectrograms, with $T=6$ timesteps for temporal encoding. 
    \item CREMA-D~\cite{cao_2014_crema_d_crowd_sourced} is a multimodal emotion recognition benchmark containing 7,442 audio-visual clips from 91 actors expressing six emotions (anger, disgust, fear, happiness, neutral, sadness). We adopt the subject-disjoint split protocol to prevent identity overfitting. All inputs use $128 \times 128$ spatial resolution for both visual frames and mel-spectrograms, with $T=6$ timesteps for temporal encoding.
    \item UrbanSound8K-AV is a multimodal extension of the UrbanSound8K audio classification~\cite{salamon_2014_dataset_and_taxonomy}, containing 8,732 labeled sound excerpts from 10 urban sound categories, including air conditioner, car horn, children playing, dog bark, drilling, engine idling, gun shot, jackhammer, siren, and street music. We pair each audio clip with synchronized video frames and apply stratified 80/10/10 train/validation/test splits to maintain class balance. All inputs use $128 \times 128$ spatial resolution for both visual frames and mel-spectrograms, with $T=6$ timesteps for temporal encoding.
\end{itemize}

\paragraph{\textbf{Methods of comparison}}
 We compare SNNergy against two categories of baselines: (1) \textit{ANN-based multimodal fusion methods} including OGM-EG~\cite{peng_2022_balanced_multimodal_learning} and PMR~\cite{fan_2023_pmr_prototypical_modal} with four fusion strategies each (sum, concatenation, FiLM, gated), representing conventional non-spiking approaches; and (2) \textit{SNN-based multimodal methods} including WeightAttention~\cite{liu_2022_event_based_multimodal}, SCA (Spiking Cross-modal Attention)~\cite{guo_2024_transformer_based_spiking}, CMCI (Cross-Modal Complementary Information)~\cite{zhou_2024_enhancing_snn_based}, S-CMRL (Spiking Cross-Modal Residual Learning)~\cite{he_2025_enhancing_audio_visual}, MISNET~\cite{liu_2025_towards_energy_efficient}, and TAAF-SNNs (Temporal Attention-guided Adaptive Fusion)~\cite{shen_2025_spiking_neural_networks}, which employ cross-modal attention in spiking networks but with quadratic complexity. For TAAF-SNNs\footnote{We use the results reported in their paper~\cite{shen_2025_spiking_neural_networks}, as the code is not publicly available.}. 

\paragraph{\textbf{Implementation Details}}
All experiments are implemented in PyTorch using the SpikingJelly~\cite{fang_2020_spikingjelly} framework for spiking neuron simulation and backpropagation through time. Models are trained on NVIDIA GPUs A100-SXM4-80GB with mixed-precision training enabled. SNNergy employs a three-stage hierarchical architecture with spatial resolution downsampling from $H/4 \to H/8 \to H/16$. For CREMA-D, we use a base embedding dimension of $d=192$ with channel progression $\{192, 384, 768\}$, while for AVE and UrbanSound8K-AV we use $d=96$ with channel progression $\{96, 192, 384\}$. Each stage contains multiple attention blocks with depths $\{1, 1, 2\}$ for Stages 1, 2, and 3 respectively. The \ac{CMQKA} mechanism in Stages 1 and 2 uses 8 attention heads, while Stage 3 employs quadratic-complexity \ac{SSA} with 8 heads for global context modeling on the coarsest feature maps.

For all datasets, input visual frames and mel-spectrograms are resized to $128 \times 128$ spatial resolution and encoded into spike trains over $T=6$ timesteps. \ac{LIF} neurons use firing threshold $\theta = 0.6$ and membrane time constant $\tau = 2.0$, with sigmoid surrogate gradient functions for backpropagation. We train all models using the AdamW optimizer with initial learning rate $\eta = 5 \times 10^{-3}$, weight decay $\lambda = 1 \times 10^{-4}$, and batch size 128. Learning rate scheduling employs a cosine annealing strategy with 5-epoch warmup. Training proceeds for 100 epochs.

\subsection{Performance Comparison Across All Datasets}
\label{subsec:performance_comparison}
\input{Table1}
\Cref{tab:performance_comparison} presents a comprehensive performance comparison of SNNergy against both ANN-based and SNN-based multimodal fusion baselines across all three benchmark datasets, and demonstrates that SNNergy achieves consistent state-of-the-art performance among \ac{SNN}-based multimodal fusion methods, validating the effectiveness of our linear-complexity \ac{CMQKA} mechanism for hierarchical cross-modal feature integration. 

\textbf{Results on the AVE Dataset:}
SNNergy achieves 72.14\% accuracy on the AVE dataset, representing a significant improvement of 1.49\% over TAAF-SNNs (70.65\%), 3.99\% over S-CMRL (68.15\%), and 4.10\% over MISNET-XL (68.04\%). The performance gap is even more substantial when compared to other spiking attention mechanisms: SNNergy outperforms SCA by 11.94\% and WeightAttention by 6.22\%, demonstrating the effectiveness of our \ac{CMQKA} mechanism for cross-modal feature integration in event-driven architectures.
When compared to \ac{ANN}-based fusion methods, it significantly outperforms all baselines while maintaining the energy efficiency advantages of spiking computation. SNNergy surpasses the best \ac{ANN} baseline (OGM-EG with Gated fusion at 64.67\%) by 7.47\%, demonstrating that energy-efficient spiking-based multimodal fusion can substantially exceed the accuracy of conventional continuous-valued approaches. The AVE dataset's diverse event categories, spanning human activities, animal sounds, musical instruments, and environmental sounds, require robust multimodal fusion to capture cross-modal semantic relationships. SNNergy's hierarchical architecture with \ac{CMQKA} at multiple scales proves particularly effective for this event classification task, as it enables progressive extraction of complementary audio-visual features from low-level acoustic-visual patterns to high-level semantic event representations.

\textbf{Results on the CREMA-D Dataset:} 
SNNergy achieves 78.38\% accuracy on CREMA-D, representing an improvement of 0.83\% over the previous best \ac{SNN} method TAAF-SNNs with concatenation fusion (77.55\%), and significant gains of 1.24\% over MISNET-XL (77.14\%), 3.16\% over MISNET-L (75.22\%), 6.79\% over S-CMRL (71.59\%), and 8.26\% over WeightAttention (70.12\%). Notably, SNNergy also surpasses the best \ac{ANN}-based fusion baseline (OGM-EG with FiLM fusion at 76.46\%) by 1.92\%, demonstrating that energy-efficient spiking-based multimodal fusion can achieve superior performance compared to conventional continuous-valued approaches.

\textbf{Results on the UrbanSound8K-AV Dataset:} 
SNNergy achieves 99.66\% accuracy on UrbanSound8K-AV, demonstrating strong performance among \ac{SNN}-based multimodal fusion methods. While \ac{ANN}-based methods achieve near-perfect accuracy on this dataset (PMR with SUM/Gated fusion at 99.77\%), SNNergy maintains competitive performance with substantially lower energy consumption through sparse, event-driven spiking computation. Among \ac{SNN} baselines, SNNergy outperforms CMCI by 1.38\%, WeightAttention by 1.19\%, S-CMRL by 1.61\%, SCA by 1.83\%, and MISNET-XL by 2.54\%, establishing state-of-the-art results among energy-efficient multimodal fusion approaches. 
The UrbanSound8K-AV dataset exhibits characteristics that differ significantly from AVE and CREMA-D: urban sound categories are highly distinctive with strong acoustic signatures (e.g., gun shot, siren, jackhammer), making the classification task less challenging compared to subtle emotion recognition or diverse event categorization. This explains the overall high performance across all methods. However, the dataset still benefits from multimodal fusion, as visual context provides complementary cues for sound source localization and disambiguation. For instance, distinguishing between different engine sounds (car vs. jackhammer) becomes more reliable when visual information about the sound source is incorporated. SNNergy's hierarchical \ac{CMQKA} mechanism enables effective audio-visual fusion even for highly distinctive sound categories, extracting cross-modal complementarity that enhances classification confidence. The smaller performance gap between \ac{SNN} and \ac{ANN} methods on this dataset (compared to AVE and CREMA-D) suggests that the task complexity is relatively low, with less room for improvement through sophisticated fusion mechanisms. 

Overall, SNNergy achieves the largest performance gains on challenging tasks requiring subtle multimodal fusion: CREMA-D emotion recognition (+0.83\% over the previous best \ac{SNN} method TAAF-SNNs) and AVE event classification (+1.49\% over TAAF-SNNs), while maintaining competitive performance on the less challenging UrbanSound8K-AV dataset. This demonstrates that our hierarchical multi-scale fusion architecture particularly benefits tasks requiring fine-grained extraction of complementary cross-modal information. 
Moreover, compared to \ac{ANN}-based baselines, SNNergy surpasses the best conventional fusion methods on CREMA-D (78.38\% vs. 76.46\% for OGM-EG with FiLM) and AVE (72.14\% vs. 64.67\% for OGM-EG with Gated), establishing that energy-efficient spiking-based multimodal fusion can exceed the accuracy of conventional continuous-valued approaches while providing substantial energy savings through event-driven computation. 
Finally, the performance advantage of SNNergy over quadratic-complexity cross-modal attention methods (S-CMRL with \ac{CCSSA}) validates our core hypothesis: linear-complexity \ac{CMQKA} enables hierarchical multi-scale fusion that captures richer cross-modal representations than single-scale, quadratic-complexity approaches. The combination of computational efficiency and fusion effectiveness positions SNNergy as a practical solution for deploying sophisticated audio-visual intelligence in resource-constrained environments where both accuracy and energy efficiency are critical.

\subsection{Ablation Study}
\label{subsec:ablation_study}
In this section, we perform additional experiments to investigate the impact of hyperparameters on the performance of SNNergy. We focus our ablation analysis on the AVE and CREMA-D datasets, as these benchmarks present more challenging multimodal fusion scenarios with greater sensitivity to architectural variations. UrbanSound8K-AV, exhibiting near-ceiling performance (99.66\% for SNNergy, 99.77\% for best \ac{ANN} baseline), offers limited dynamic range for meaningful ablation analysis due to the highly distinctive acoustic signatures of urban sound categories. 

\paragraph{\textbf{Temporal Encoding Depth Analysis}}
\begin{table}[t]
    \centering
    \caption{Audio-visual timestep analysis on AVE and CREMA-D datasets.}
    \label{tab:ablation_timesteps}
    \begin{tabular}{llcc}
        \hline
        \textbf{Dataset} & \textbf{Timestep} & \textbf{Top-1 Accuracy (\%)} & \textbf{Test Loss} \\
        \hline
        \multirow{4}{*}{AVE} & 1                 & 64.18                        & 1.2972             \\
                             & 2                 & 68.66                        & 1.2867             \\
                             & 4                 & 69.99                        & 1.2108             \\
                             & 6                 & 72.14                        & 1.1667             \\
        \hline
        \multirow{4}{*}{CREMA-D} & 1                 & 72.63                        & 0.8299             \\
                                 & 2                 & 73.71                        & 0.7873             \\
                                 & 4                 & 75.75                        & 0.7504             \\
                                 & 6                 & 78.38                        & 0.7302             \\
        \hline
    \end{tabular}
\end{table}

To investigate the impact of temporal depth on multimodal fusion performance, we vary the number of timesteps $T \in \{1, 2, 4, 6\}$ while keeping all other architectural configurations fixed. As shown in \Cref{tab:ablation_timesteps}, SNNergy demonstrates consistent performance improvements as temporal resolution increases across both benchmarks. On the AVE dataset, accuracy rises from 64.18\% at $T=1$ to 72.14\% at $T=6$, corresponding to a substantial gain of 7.96\%. A similar trend is observed on the CREMA-D dataset, where accuracy improves from 72.63\% to 78.38\% (+5.75\%). This consistent upward trend indicates that increased temporal encoding depth enables more precise spike-timing-dependent feature representations, which are critical for capturing fine-grained cross-modal synchronization in audio-visual events and emotions.
The test loss exhibits a corresponding monotonic decrease on both datasets (AVE: 1.2972 to 1.1667; CREMA-D: 0.8299 to 0.7302), further confirming improved model convergence and generalization with higher temporal resolution. Notably, the performance gains show diminishing returns  beyond $T=4$, where reduced marginal improvements are observed at higher timestep counts. Based on this analysis, we adopt $T=6$ as the default configuration for all subsequent experiments, balancing fusion accuracy with computational cost.

\paragraph{\textbf{Component Contribution Analysis}}
\label{tab:component_contribution_analysis}

\begin{table}[t]
\centering
\caption{Impact and contributions of spatial and temporal components in SNNergy, tested on the AVE and CREMA-D datasets.}
\label{tab:ablation}
\begin{tabular}{llcc}
        \hline
        \textbf{Dataset} & \textbf{Component} & \textbf{Top-1 Accuracy (\%)} & \textbf{Test Loss} \\
        \hline
        \multirow{3}{*}{AVE} & Spatial-only                & 63.68  & 1.3691             \\
                             & Temporal-only               & 70.15 & 1.1849             \\
                             & Spatiotemporal              & 72.14 & 1.1667            \\
        \hline
        \multirow{3}{*}{CREMA-D} & Spatial-only                & 77.78  & 0.7422             \\
                                 & Temporal-only               & 77.24 & 0.7749             \\
                                 & Spatiotemporal              & 78.38 & 0.7302               \\
        \hline
    \end{tabular}

\end{table}

To understand the contribution of different components to SNNergy's overall performance, we test three variants of our model: spatial-only, temporal-only, and spatiotemporal. 
As shown in Table~\ref{tab:ablation}, the results vary across datasets, highlighting the different characteristics of audio-visual information in each task. On the AVE dataset, the temporal-only model (70.15\%) significantly outperforms the spatial-only model (63.68\%), indicating that temporal dynamics are more discriminative for event classification. However, on the CREMA-D dataset, the spatial-only model (77.78\%) achieves higher accuracy than the temporal-only model (77.24\%), suggesting that spatial facial expressions and visual cues are particularly strong indicators for emotion recognition. 
In both cases, combining both spatial and temporal components in the spatiotemporal model yields the best performance, achieving 72.14\% on AVE and 78.38\% on CREMA-D. This consistent improvement demonstrates that spatial and temporal features provide complementary information across different audio-visual tasks. The integration of both pathways through our \ac{CMQKA} mechanism effectively captures these multifaceted relationships, leading to superior classification performance. 
Table~\ref{tab:ablation} summarizes the results of our ablation study, clearly showing the incremental benefit of incorporating both spatial and temporal modalities in our architecture.

\paragraph{\textbf{Spike Firing Rate Analysis}}
\begin{table}[tb]
    \centering
    \caption{Spike firing rates in CMQKA attention blocks across stages on CREMA-D and AVE datasets. Values are averaged across video/audio modalities and spatial/temporal paths. Results show attention masks (Attn) have higher firing rates than Q/K neurons, indicating their role as learned gating mechanisms.}
    \label{tab:firing_rates}
    \begin{tabular}{l|ccc|ccc}
        \hline
        & \multicolumn{3}{c|}{\textbf{CREMA-D}} & \multicolumn{3}{c}{\textbf{AVE}} \\
        \textbf{Component} & \textbf{CM1} & \textbf{CM2} & \textbf{SSA} & \textbf{CM1} & \textbf{CM2} & \textbf{SSA} \\
        \hline
        \multicolumn{7}{c}{\textbf{QK Attention Blocks}} \\
        \hline
        Q    & 0.1304 & 0.1396 & 0.1285 & 0.0896 & 0.1027 & 0.1470 \\
        K    & 0.1423 & 0.1348 & 0.1077 & 0.1258 & 0.1154 & 0.1278 \\
        Attn & 0.4355 & 0.3842 & 0.2156 & 0.3103 & 0.3247 & 0.1890 \\
        \hline
        \multicolumn{7}{c}{\textbf{MLP Feed-Forward Blocks}} \\
        \hline
        Layer1 & 0.1158 & 0.1076 & 0.0814 & 0.0840 & 0.0981 & 0.0832 \\
        Layer2 & 0.0987 & 0.0775 & 0.0671 & 0.0769 & 0.0617 & 0.0678 \\
        \hline
    \end{tabular}
\end{table}

As shown in \Cref{tab:firing_rates}, the firing rate distribution across both CREMA-D and AVE datasets reveals that the Query and Key neurons in \ac{CMQKA} maintain remarkably low firing rates of approximately 9--15\% across all stages and datasets, demonstrating the sparse, event-driven nature of cross-modal attention computation. This sparsity directly translates to energy savings, as only a small fraction of neurons actively consume power per timestep, compared to 100\% activation in conventional \ac{ANN}-based attention mechanisms.

The attention mask neurons (Attn) exhibit moderately higher firing rates, ranging from 19\% to 44\%, which is approximately 2--4$\times$ higher than those of the Query/Key neurons. This elevated firing rate indicates that the attention masks function as active learned gating mechanisms rather than passive filtering operations. The threshold of 0.6 in the attention \ac{LIF} neurons creates a binary selection gate that adaptively determines which cross-modal tokens are relevant for fusion. Notably, the firing rates exhibit a hierarchical pattern across stages: on CREMA-D, CM1 (43.55\%) $>$ CM2 (38.42\%) $>$ SSA (21.56\%), and similarly on AVE, CM1 (31.03\%) $\approx$ CM2 (32.47\%) $>$ SSA (18.90\%). This progressive reduction in later stages suggests that early stages perform broad cross-modal feature extraction with higher token selection, while later stages focus on selective refinement of discriminative multimodal features. The lower SSA firing rates reflect the coarser spatial resolution at Stage 3 ($H/16 \times W/16$), where fewer but more semantically meaningful tokens require attention. 
CREMA-D exhibits higher attention firing rates than AVE across most stages, reflecting the different task characteristics: emotion recognition requires more intensive cross-modal feature extraction from facial expressions and speech patterns, whereas event classification relies on more selective temporal cues.

The \ac{MLP} feed-forward blocks maintain low firing rates of approximately 6--12\% across all stages and layers, ensuring balanced feature transformation without excessive spike activity. The consistency of MLP firing rates across Layer1 (expansion) and Layer2 (projection) indicates stable gradient flow and effective residual learning throughout the hierarchical architecture. Importantly, the similar or lower firing rates between MLP layers and Query/Key neurons demonstrate that SNNergy achieves computational balance: cross-modal attention mechanisms do not introduce disproportionate spike activity compared to standard feed-forward processing.

From an energy efficiency perspective, these firing rate patterns validate SNNergy's design principles. The low firing rates in Query/Key projections (9--15\%) and MLP layers (6--12\%) ensure that the majority of computations remain sparse and event-driven, directly reducing synaptic operations and power consumption. The moderate firing rates in attention masks (19--44\%) are justified by their critical role as adaptive gating mechanisms that enable effective cross-modal fusion. Since attention masks operate on channel-aggregated representations rather than full token-to-token matrices, their computational overhead remains manageable. The hierarchical decrease in attention firing rates further contributes to efficiency, as later stages process fewer tokens with more focused selection.

\subsection{Computational Analysis of the CMQKA Mechanism}
\label{subsec:cmqka_analysis}


\paragraph{\textbf{Fusion Complexity Analysis}}
\begin{table}[t]
\centering
\caption{Fusion Complexity Comparison between \ac{CMQKA} ($\mathcal{O}(NC^2)$) and \ac{SSA} ($\mathcal{O}(N^2C)$) for standard downsampling ($4\times, 8\times, 16\times$) with input resolution $128 \times 128$. Bold indicates the more efficient mechanism at each stage, justifying the hybrid design.}
\label{tab:complexity_standard}
\resizebox{0.9\textwidth}{!}{
\begin{tabular}{@{}lccccc@{}}
\toprule
\multirow{2}{*}{Stage} & \multirow{2}{*}{$N$} & \multicolumn{2}{c}{$C$: $192 \rightarrow 384 \rightarrow 768$} & \multicolumn{2}{c}{$C$: $96 \rightarrow 192 \rightarrow 384$} \\
\cmidrule(lr){3-4} \cmidrule(lr){5-6}
& & CMQKA & SSA & CMQKA & SSA \\
\midrule
1 ($32\!\times\!32$) & 1024 & \textbf{37.7} & 201.3 & \textbf{9.4} & 100.7 \\
2 ($16\!\times\!16$) & 256  & 37.7 & \textbf{25.2} & \textbf{9.4} &  12.6 \\
3 ($8\!\times\!8$)   & 64   & 37.7          & \textbf{3.1} & 9.4 & \textbf{1.6} \\
\midrule
\multicolumn{2}{@{}l}{Hybrid Total (M ops)} & \multicolumn{2}{c}{\textbf{78.5}} & \multicolumn{2}{c}{\textbf{20.4}} \\
\multicolumn{2}{@{}l}{All-CMQKA Total (M ops)} & \multicolumn{2}{c}{113.1} & \multicolumn{2}{c}{28.2} \\
\multicolumn{2}{@{}l}{All-SSA Total (M ops)} & \multicolumn{2}{c}{229.6} & \multicolumn{2}{c}{114.9} \\
\bottomrule
\end{tabular}
}
\end{table}

To optimize the trade-off between local feature granularity and global context capture, SNNergy employs a hybrid fusion strategy governed by the crossover relationship between the number of spatial tokens $N$ and the channel dimension $C$.
The computational complexity of the proposed \ac{CMQKA} mechanism is $\mathcal{O}(NC^2)$, whereas standard \ac{SSA} requires $\mathcal{O}(N^2C)$. Consequently, the efficiency advantage of \ac{CMQKA} is maximized in early hierarchical stages where spatial resolution is high ($N \gg C$), while \ac{SSA} becomes competitive only in the deepest stages where aggressive downsampling results in feature maps where $C > N$.\Cref{tab:complexity_standard} presents a fusion complexity comparison across the three hierarchical stages of SNNergy. In Stage 1, where the fusion of fine-grained audio-visual features occurs at $32 \times 32$ resolution ($N{=}1024$), \ac{CMQKA} drastically reduces the computational load compared to \ac{SSA} (e.g., 9.4M vs. 100.7M operations at $C{=}96$). This efficiency enables SNNergy to perform dense cross-modal fusion at high resolutions, a capability often discarded in other frameworks to save cost.
As the hierarchy progresses to Stage 3, the spatial dimension contracts ($N{=}64$) while the semantic channel dimension expands ($C{=}384$ or $C{=}768$). Here, the architecture strategically transitions to \ac{SSA} to capture global multimodal context, leveraging the channel-linear properties of \ac{SSA} when $C \gg N$.
This balanced complexity design results in a total operation count of 78.5M for the hybrid architecture (at base dim 192), which is $1.4\times$ more efficient than an all-\ac{CMQKA} design and $2.9\times$ more efficient than an all-\ac{SSA} design.

\begin{figure}[tb]
    \centering
    \includegraphics[width=0.8\linewidth]{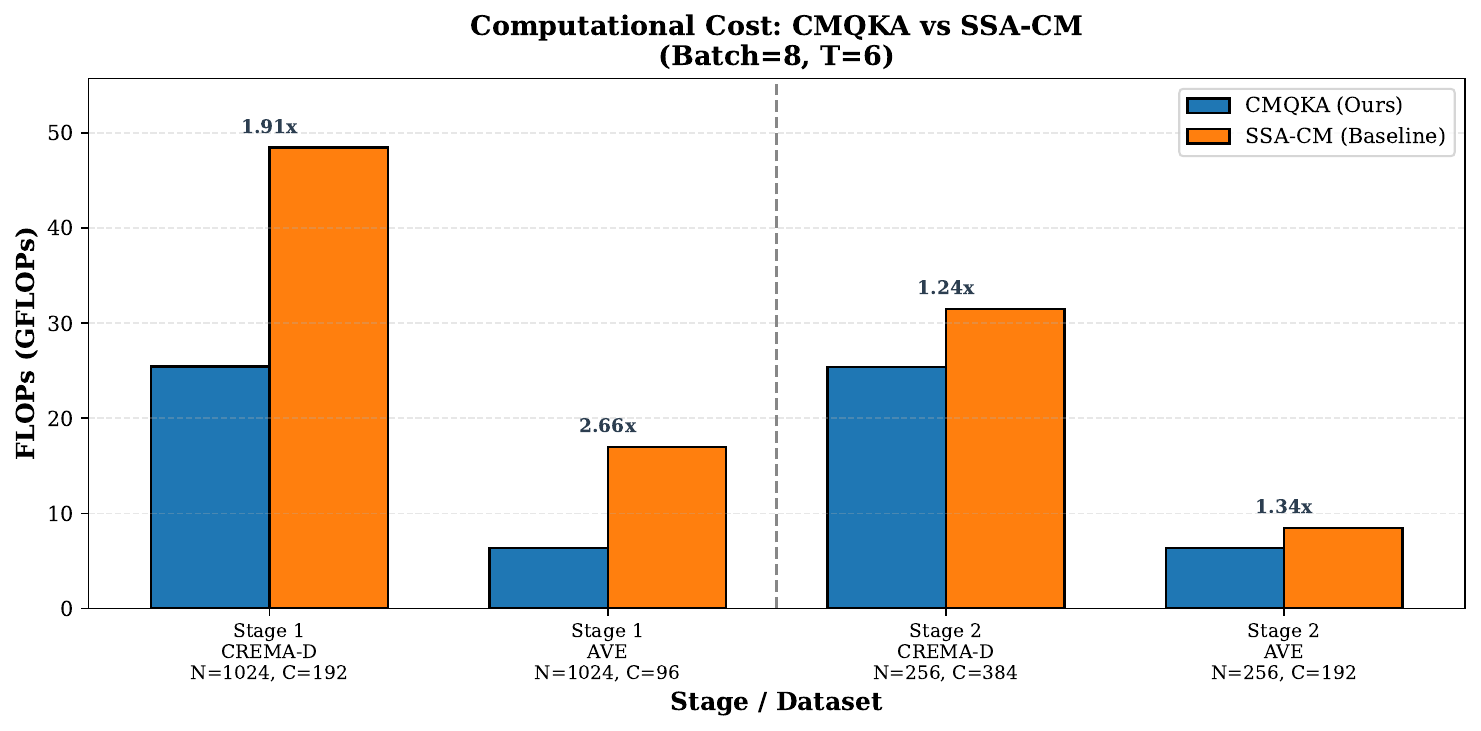}
    \caption{FLOPs comparison across stages 1 and 2 on both CREMA-D and AVE datasets, demonstrating the computational efficiency of \ac{CMQKA}'s linear-complexity fusion mechanism. The gap between quadratic and linear attention grows with spatial resolution, validating the scalability benefits of our approach across diverse audio-visual benchmarks.}
    \label{fig:flops_comparison}
\end{figure}

\paragraph{\textbf{Runtime Performance}}
We conduct empirical measurements comparing computational efficiency across stages 1 and 2 of the hierarchical architecture on both CREMA-D and AVE datasets. \Cref{fig:flops_comparison} presents the floating-point operations comparison between different architectural configurations across both benchmarks. The results demonstrate that \ac{CMQKA}'s linear complexity translates directly into reduced computational requirements: as the spatial resolution increases from Stage 2 to Stage 1, the FLOPs gap between quadratic-complexity attention (e.g., \ac{CCSSA}) and our linear-complexity \ac{CMQKA} widens substantially. This scaling behavior validates the theoretical complexity analysis, where the reduction factor of $O(N/C)$ becomes increasingly significant at higher resolutions. The FLOPs savings are particularly pronounced in Stage 1, where finer-grained spatial tokens ($H \times W$) require more extensive attention computation.

\begin{figure}[tb]
    \centering
    \includegraphics[width=0.8\linewidth]{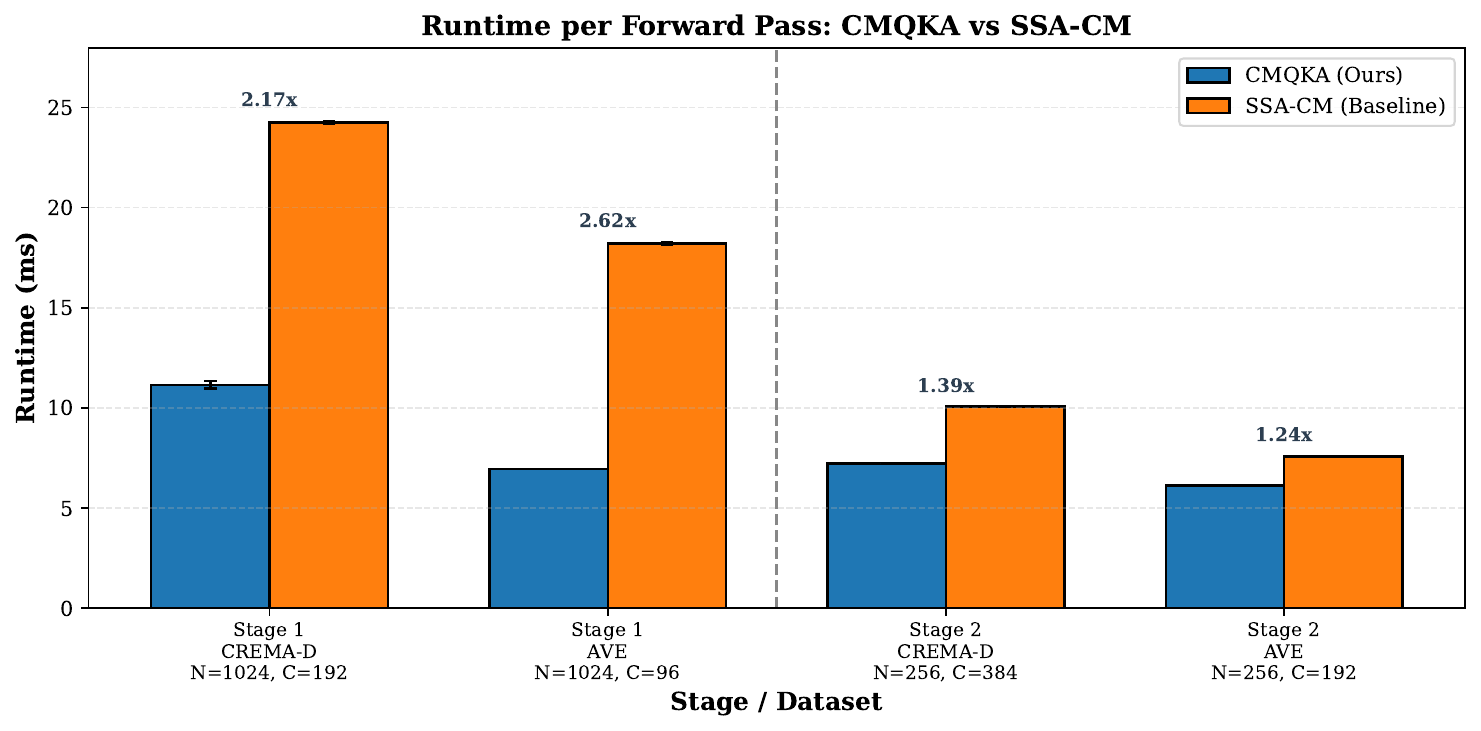}
    \caption{Runtime comparison across stages 1 and 2 on both CREMA-D and AVE datasets, showing the inference speed advantage of linear-complexity \ac{CMQKA}. Faster execution enables real-time multimodal processing for latency-sensitive applications.}
    \label{fig:runtime_comparison}
\end{figure}

The theoretical reduction in operation count translates directly to faster cross-modal inference, a critical requirement for real-time multimodal systems. \Cref{fig:runtime_comparison} illustrates the runtime performance of the fusion layers across Stages 1 and 2. \ac{CMQKA} demonstrates significantly lower latency compared to quadratic alternatives, particularly in Stage 1 where the high token count imposes a severe penalty on conventional attention mechanisms. This runtime efficiency ensures that the integration of audio and visual streams does not become the bottleneck in the inference pipeline, facilitating low-latency decision-making in time-sensitive applications.

\paragraph{\textbf{Memory Analysis}}
The linear \ac{CMQKA} mechanism reduces memory complexity from $O(N^2)$ to $O(N)$ by maintaining only the attention vector $\mathbf{A}_t \in \{0,1\}^{N}$ rather than the full $N \times N$ attention matrix required by quadratic-complexity approaches such as \ac{CCSSA}.

\begin{figure}[tb]
    \centering
    \includegraphics[width=0.8\linewidth]{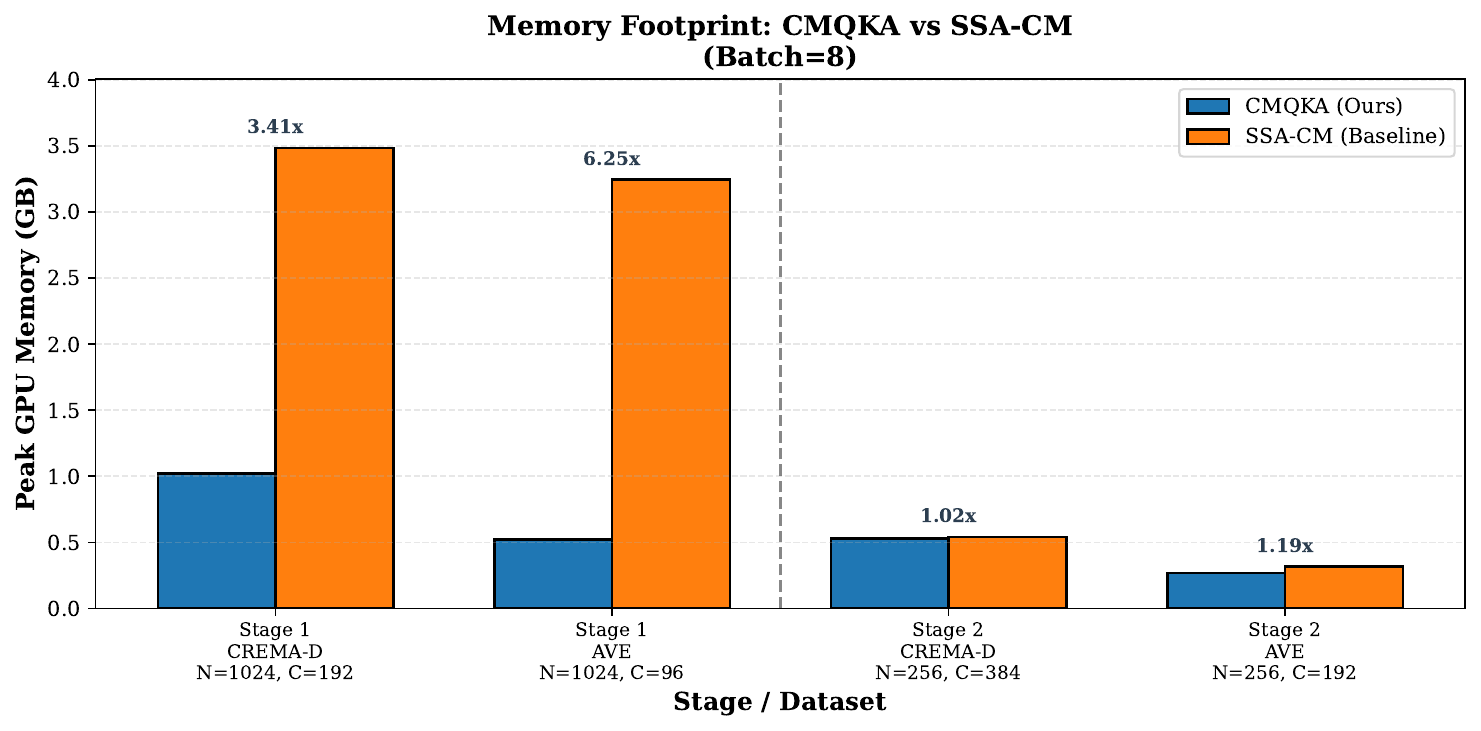}
    \caption{Memory consumption comparison across stages 1 and 2 on both CREMA-D and AVE datasets. \ac{CMQKA}'s linear memory complexity compared to quadratic attention notably reduces the memory footprint, particularly critical for neuromorphic deployment.}
    \label{fig:memory_comparison}
\end{figure}

\Cref{fig:memory_comparison} illustrates memory consumption patterns across stages 1 and 2 of the hierarchical architecture on both datasets. The results demonstrate that \ac{CMQKA} maintains a substantially lower memory footprint compared to quadratic-complexity attention mechanisms. This reduction becomes especially important in hierarchical architectures where multiple fusion stages operate in parallel: maintaining quadratic-complexity attention across all stages would require excessive memory allocation, limiting the feasibility of multi-scale processing. SNNergy's linear memory scaling enables efficient hierarchical fusion while remaining compatible with the memory constraints of edge deployment scenarios.

\begin{figure}[tb]
    \centering
    \includegraphics[width=0.8\linewidth]{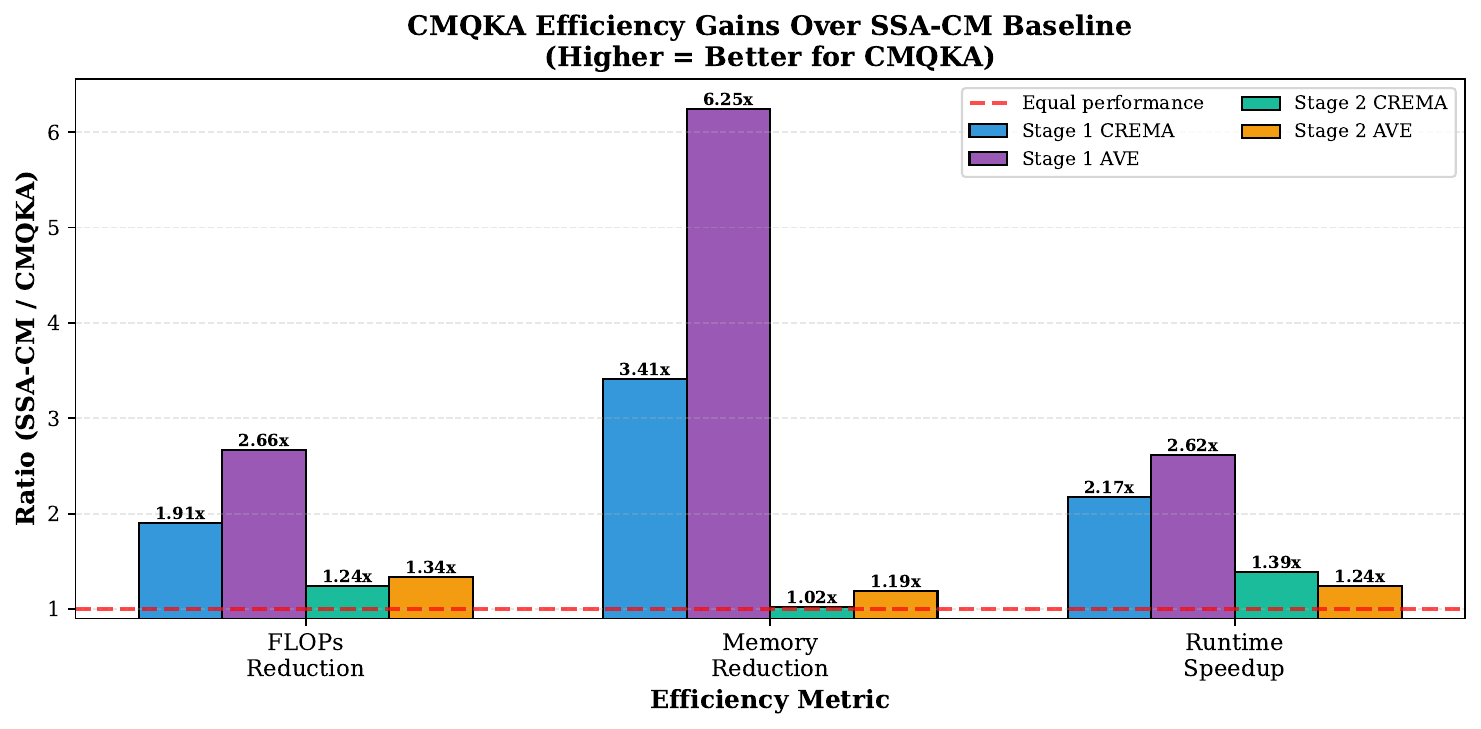}
    \caption{Comprehensive efficiency analysis integrating FLOPs, memory, and runtime metrics across stages 1 and 2 on both CREMA-D and AVE datasets. The consistent efficiency advantages across all dimensions and datasets demonstrate \ac{CMQKA}'s effectiveness as a practical solution for energy-efficient hierarchical multimodal fusion.}
    \label{fig:efficiency_analysis}
\end{figure}

\paragraph{\textbf{Comprehensive Efficiency Summary.}} \Cref{fig:efficiency_analysis} provides an integrated view of the efficiency achieved by \ac{CMQKA} across all measured dimensions: FLOPs, memory, and runtime, evaluated on both CREMA-D and AVE. The analysis indicates that linear-complexity attention provides consistent efficiency improvements that compound across the hierarchical architecture and generalize across different audio-visual benchmarks. Importantly, these empirical measurements validate our theoretical complexity analysis: the practical benefits of $O(N)$ versus $O(N^2)$ scaling are not merely asymptotic theoretical advantages but translate into tangible computational savings even at moderate resolutions ($128 \times 128$ inputs). The efficiency gains become more pronounced in hierarchical multi-stage architectures where fusion operations are repeated at multiple scales, making \ac{CMQKA} an enabling technology for sophisticated multimodal fusion in energy-constrained environments.

These efficiency measurements provide concrete evidence that SNNergy's design principles, linear-complexity attention, hierarchical multi-scale fusion, and event-driven spiking computation, collectively enable the practical deployment of sophisticated audio-visual intelligence on resource-constrained platforms. The demonstrated efficiency gains in FLOPs, memory, and runtime, combined with the competitive accuracy results presented earlier, position SNNergy as a viable solution for bringing multimodal fusion capabilities to edge devices, mobile platforms, and neuromorphic processors.

\section{Conclusion}
\label{sec:conclusion}

In this paper, we presented SNNergy, a novel energy-efficient multimodal fusion framework for audio-visual intelligence that addresses the fundamental challenge of efficient cross-modal feature integration in resource-constrained environments. Our key contribution is the \ac{CMQKA} fusion mechanism, which achieves linear computational complexity for multimodal fusion through query-key attention while maintaining competitive recognition performance across diverse audio-visual tasks. 
The hierarchical architecture of SNNergy progressively integrates multimodal features across three stages with increasing semantic abstraction, enabled by the \ac{SPDS} module that preserves spike information during spatial downsampling. Experimental results on CREMA-D, AVE, and UrbanSound8K-AV datasets demonstrate that SNNergy outperforms existing \ac{SNN}-based multimodal methods while maintaining the energy efficiency advantages of event-driven spiking computation. 

Our work demonstrates that efficient multimodal fusion is achievable in spiking neural networks without sacrificing performance, opening new possibilities for deploying advanced audio-visual intelligence on edge devices and neuromorphic hardware. The linear complexity of \ac{CMQKA} makes SNNergy particularly suitable for real-time applications where both computational efficiency and energy constraints are critical.
Future work will explore extending SNNergy to other multimodal domains beyond audio-visual processing, investigating adaptive temporal encoding strategies to further optimize energy consumption, and deploying the framework on neuromorphic hardware platforms to validate end-to-end energy efficiency in practical scenarios.

\bibliographystyle{elsarticle-num}
\bibliography{root}

\end{document}

%% file: acros.tex
\DeclareAcronym{SNN}{short = SNN, long = Spiking Neural Network}
\DeclareAcronym{ANN}{short = ANN, long = Artificial Neural Network}
\DeclareAcronym{BPTT}{short = BPTT, long = Backpropagation Through Time}
\DeclareAcronym{STBP}{short = STBP, long = Spatio-temporal Backpropagation}
\DeclareAcronym{OTTT}{short = OTTT, long = Online Training Through Time}
\DeclareAcronym{SLTT}{short = SLTT, long = Spatial Learning Through Time}
\DeclareAcronym{SSA}{short = SSA, long = Spiking Self-Attention}
\DeclareAcronym{SDSA}{short = SDSA, long = Spike-Driven Self-Attention}
\DeclareAcronym{QKTA}{short = QKTA, long = Q-K Token Attention}
\DeclareAcronym{QKCA}{short = QKCA, long = Q-K Channel Attention}
\DeclareAcronym{SPEDS}{short = SPEDS, long = Spiking Patch Embedding with Deformed Shortcut}
\DeclareAcronym{SPDS}{short = SPDS, long = Spiking Patch Downsampling with Shortcut}
\DeclareAcronym{SCMRL}{short = SCMRL, long = Semantic-Alignment Cross-Modal Residual Learning}
\DeclareAcronym{CCSSA}{short = CCSSA, long = Cross-modal Complementary Spatio-temporal Spiking Attention}
\DeclareAcronym{SCSA}{short = SCSA, long = Spatial Complementary Spiking Attention}
\DeclareAcronym{TCSA}{short = TCSA, long = Temporal Complementary Spiking Attention}
\DeclareAcronym{SAO}{short = SAO, long = Semantic Alignment Optimization}
\DeclareAcronym{RNN}{short = RNN, long = Recurrent Neural Network}
\DeclareAcronym{MAC}{short = MAC, long = Multiply-Accumulate}
\DeclareAcronym{AC}{short = AC, long = Accumulate}
\DeclareAcronym{CMQKA}{short = CMQKA, long = Cross Modal Q-K Attention}
\DeclareAcronym{LIF}{short = LIF, long = Leaky Integrate-and-Fire}
\DeclareAcronym{MLP}{short = MLP, long = Multi-Layer Perceptron}
\DeclareAcronym{SN}{short = SN, long = Spiking Neuron}
\DeclareAcronym{AMP}{short = AMP, long = Automatic Mixed Precision}

%% file: Table1.tex
\begin{table}[tb]
    \centering
    \caption{Performance comparison on CREMA-D, AVE, and UrbanSound8K-AV datasets in terms of accuracy. SNNergy achieves state-of-the-art results among SNN-based methods across all benchmarks. $^\dagger$ indicates reported results from the respective paper.}
    \label{tab:performance_comparison}
    \resizebox{\textwidth}{!}{%
    \begin{tabular}{llcccr}
        \hline
        \textbf{Model} & \textbf{Method} & \textbf{Fusion} & \textbf{CREMA-D} & \textbf{AVE} & \textbf{Us8k-AV} \\
        \hline
        \multirow{8}{*}{ANN} 
            & \multirow{4}{*}{OGM-EG~\cite{peng_2022_balanced_multimodal_learning}} 
                & SUM    & 74.80 & 62.68 & 98.97 \\
            &   & Concat & 74.02 & 64.42 & 98.97 \\
            &   & Film   & 76.46 & 61.44 & 98.74 \\
            &   & Gated  & 72.67 & 64.67 & 99.08 \\
        \cline{2-6}
            & \multirow{4}{*}{PMR~\cite{fan_2023_pmr_prototypical_modal}} 
                & SUM    & 74.88 & 63.43 & 99.77 \\
            &   & Concat & 69.51 & 62.69 & 99.66 \\
            &   & Film   & 75.73 & 61.69 & 99.66 \\
            &   & Gated  & 70.73 & 48.26 & 99.77 \\
        \hline

        \multirow{9}{*}{SNN}
            & WeightAttention~\cite{liu_2022_event_based_multimodal} 
                & --- & 70.12 & 65.92 & 98.17 \\
        \cline{2-6}
            & SCA~\cite{guo_2024_transformer_based_spiking} 
                & --- & 67.80 & 60.20 & 97.83 \\
        \cline{2-6}
            & CMCI~\cite{zhou_2024_enhancing_snn_based} 
                & --- & 71.59 & 67.91 & 98.28 \\
        \cline{2-6}
            & S-CMRL~\cite{he_2025_enhancing_audio_visual} 
                & --- & 71.59 & 68.15 & 98.05 \\
        \cline{2-6}
            & MISNET-L~\cite{liu_2025_towards_energy_efficient} 
                & --- & 75.22 & 67.24 & 95.23 \\
            & MISNET-XL~\cite{liu_2025_towards_energy_efficient} 
                & --- & 77.14 & 68.04 & 97.12 \\
        \cline{2-6}
            & TAAF-SNNs$^\dagger$~\cite{shen_2025_spiking_neural_networks} 
                & Concat & 77.55 & 70.65 & --- \\
            & TAAF-SNNs$^\dagger$~\cite{shen_2025_spiking_neural_networks} 
                & Sum    & 76.75 & 68.91 & --- \\
        \cline{2-6}
            & \bf{SNNergy (ours)} 
                & --- & \bf{78.38} & \bf{72.14} & \bf{99.66} \\
        \hline
    \end{tabular}
    }
\end{table}